\documentclass[aps,pra,twocolumn,superscriptaddress,nofootinbib,longbibliography]{revtex4-2}
\usepackage{amsmath,amssymb,graphicx,xspace,booktabs,siunitx}
\usepackage[colorlinks=true,linkcolor=blue,citecolor=blue,urlcolor=blue]{hyperref}
\newtheorem{proposition}{Proposition}

\newcommand{\posBackend}{ibm\_fez\xspace}
\newcommand{\posQubit}{46\xspace}

\newcommand{\nuMeasQs}{1.72\xspace}
\newcommand{\nuErrQs}{0.15\xspace}

\newcommand{\nuMeasMid}{1.51\xspace}
\newcommand{\nuErrMid}{0.08\xspace}

\newcommand{\nuMeasFast}{1.00\xspace}
\newcommand{\nuErrFast}{0.10\xspace}

\newcommand{\posMaxZ}{1.61\xspace}
\newcommand{\wvarRepQsDD}{4.02\xspace}

\newcommand{\nullNu}{1.05\xspace}
\newcommand{\nullNuErr}{0.05\xspace}
\newcommand{\nullQubit}{46\xspace}
\newcommand{\auditChiMin}{15.81\xspace}
\newcommand{\auditChiMax}{18.97\xspace}
\newcommand{\auditBallCut}{12.59\xspace}
\newcommand{\cycleResetError}{4.66\times 10^{-15}\xspace}
\newcommand{\cycleFormulaError}{1.39\times 10^{-16}\xspace}
\newcommand{\cycleTwirlError}{3.33\times 10^{-16}\xspace}
\newcommand{\cycleBlocks}{512\xspace}
\newcommand{\cycleTrials}{200\xspace}
\newcommand{\cycleTruth}{0.1704\xspace}
\newcommand{\cycleRtnDetect}{100.0\xspace}
\newcommand{\cycleRtnCoverage}{92.5\xspace}
\newcommand{\cycleRtnFinite}{100.0\xspace}
\newcommand{\cycleResetDetect}{4.5\xspace}
\newcommand{\cycleResetCoverage}{95.5\xspace}
\newcommand{\cycleResetFinite}{100.0\xspace}
\newcommand{\cycleStaticDetect}{6.5\xspace}
\newcommand{\cycleStaticCoverage}{93.5\xspace}
\newcommand{\cycleStaticFinite}{100.0\xspace}
\newcommand{\cycleWhiteDetect}{5.5\xspace}
\newcommand{\cycleWhiteCoverage}{94.5\xspace}
\newcommand{\cycleWhiteFinite}{100.0\xspace}

\newcommand{\cycleFirstPrimary}{0.254\xspace}
\newcommand{\cycleFirstLo}{0.177\xspace}
\newcommand{\cycleFirstHi}{0.331\xspace}
\newcommand{\cycleSecondPrimary}{0.211\xspace}
\newcommand{\cycleSecondLo}{0.136\xspace}
\newcommand{\cycleSecondHi}{0.285\xspace}

\newcommand{\devTwo}{ibm\_marrakesh\xspace}

\newcommand{\devTwoTopQubit}{154\xspace}
\newcommand{\devTwoTopNu}{2.63\xspace}
\newcommand{\devTwoTopErr}{0.54\xspace}

\newcommand{\devTwoConfNu}{1.56\xspace}
\newcommand{\devTwoConfErr}{0.28\xspace}

\newcommand{\totalQubits}{16\xspace}

\newcommand{\ASF}{\mathrm{ASF}}
\newcommand{\EPC}{\mathrm{EPC}}
\newcommand{\tc}{\tau_c}
\newcommand{\dd}{\delta}
\newcommand{\Wrep}{W_{\mathrm{var}}^{\mathrm{rep}}}
\newcommand{\Cov}{\operatorname{Cov}}

\begin{document}
\title{Phase-cycled randomized benchmarking of quantum processors:\\
recovering hidden classical noise correlations}

\author{Mirza Samad Ahmed Baig}
\email{Mirza@fandaqah.com}
\affiliation{Fandaqah, Al Khobar, Saudi Arabia}
\author{Syeda Anshrah Gillani}
\email{syeda.gilani@stud.uni-heidelberg.de}
\affiliation{Heidelberg University, Heidelberg, Germany}
\author{Abdul Akbar Khan}
\email{Akbar.khan@Argaam.com}
\affiliation{Argaam, Riyadh, Saudi Arabia}
\author{Muhammad Omer Khan}
\email{Omer.khan@fortanixor.com}
\affiliation{Fortanixor, United Arab Emirates}

\begin{abstract}
Randomized benchmarking can hide classical temporal correlations because its
Clifford-twirled response is even in the noise phase. For a stationary symmetric
telegraph fluctuator, we show that continuous evolution and independent stationary
resets at slot boundaries yield identical mean responses for arbitrary fixed idle
modulations. We construct an eight-setting phase-cycle measurement of the connected
sine-phase covariance under ideal Clifford twirling and classical idle dephasing.
This observable vanishes for independent slot noise and fixed detuning without a
weak-phase or Gaussian approximation. A closed telegraph response, independent
circuit calculations and 800 simulation trials validate the construction and
quantify the empirical coverage of a paired bootstrap estimator. A separate
conservative confidence set states its finite-sample assumptions. Two acquisitions
on an IBM processor compare engineered shared-sign and independently reset phases
with identical marginals. Their primary contrasts are
$\cycleFirstPrimary$ and $\cycleSecondPrimary$, with empirical 95\% intervals
$[\cycleFirstLo,\cycleFirstHi]$ and $[\cycleSecondLo,\cycleSecondHi]$,
respectively; all negative-control intervals include zero. At equal shot and
sensing-window budgets, an ideal Ramsey/echo estimator is more precise in every
tested class. The result supplies an explicit connection between a benchmarking
identifiability limitation and a controlled correlation measurement. No native
or quantum-memory detection is claimed.
\end{abstract}
\maketitle
\raggedbottom

\section{Introduction}
Randomized benchmarking (RB) estimates a compact description of gate performance
by averaging random circuits~\cite{Emerson2005,Knill2008,Magesan2011}. The single
exponential associated with gate-independent memoryless noise is attractive
experimentally, but observing that shape does not establish the noise assumptions.
Classical environments can preserve information across gates while producing the
same averaged decay~\cite{Ball2016,Qi2021,BlindSpots2026}. This distinction matters
because a fitted average error is not itself a worst-case error certificate.

Idle-duration scans are an established way to probe this problem. RB Ramsey and
RB echo already measured dephasing versus idle duration, resolved telegraph noise,
and separated linear and quadratic error contributions on superconducting
qubits~\cite{OMalley2015}. Engineered-noise RB experiments also measured and
suppressed error correlations, with theoretical predictions tested on trapped
ions~\cite{Edmunds2020}. Accordingly, neither insertion of an idle window nor
calibration against injected correlated noise is claimed here as a new principle.

Our question is narrower: which feature of the Clifford average hides a symmetric
fluctuator, and can a small change of the measured observable recover the information
it removes? We answer by separating two tasks that an idle scan can otherwise
conflate. Measuring the duration dependence of the phase distribution within a slot
is a spectroscopy task. Distinguishing continuous environmental evolution from
independent resets between slots is an identifiability task. Even an ideal, arbitrarily
dense gap scan need not solve the second task.

The contribution is a constructive pair of results. First, we establish an exact
reset equivalence for a symmetric telegraph environment, including unequal idle
durations. Second, controlled phase cycling converts the even Clifford response into
a connected odd-phase correlator. The latter responds to inter-slot dependence while
rejecting a fixed detuning and a reset model with identical marginal distributions.
Both statements concern an explicitly specified dephasing model; neither is a
universal classifier of classical versus quantum non-Markovianity.

Existing analyses explain non-exponential RB under correlated dephasing
\cite{Fogarty2015,Qi2021}, temporal-noise effects with finite-duration
gates~\cite{Brillant2025}, and learning correlated dynamics from RB data
\cite{Yang2022,Zhang2025}. Multi-exponential estimation has a general framework
\cite{HelsenFramework2022}, and finite-sample RB statistics have substantial prior
development~\cite{Wallman2014}. Sine-phase correlations already appear in
Ramsey noise spectroscopy~\cite{Yan2012}; phase cycling and correlation
spectroscopy are established beyond RB~\cite{Meinel2022}, including the recent
RESOLUTE protocol~\cite{Zohar2026}. Our proposed contribution is the explicit
Clifford-twirled implementation of a connected sine covariance, its constructive
reset-equivalence motivation, and its matched-marginal controlled test. Neither
the existence of RB blindness nor sine-correlation spectroscopy itself is new.
Section~\ref{sec:ramsey} compares an ideal Ramsey/echo construction at equal
circuit-shot and sensing-window budgets.

\section{What a gap scan can identify}
\label{sec:gap}
Consider ideal independent single-qubit Cliffords separated by dephasing slots.
Only the Cliffords enter the final inverse. A slot accumulates a classical phase
$\phi_k$, independent of the chosen Cliffords. Averaging the Clifford sequence gives
\begin{equation}
\label{eq:twirl}
 p(\phi)=\frac{1+2\cos\phi}{3},\qquad
 Z_m=\mathbb{E}\!\left[\prod_{k=1}^{m}p(\phi_k)\right],
\end{equation}
where $\ASF(m)$ is the average sequence fidelity (survival probability) and
$Z_m=2\ASF(m)-1$ for ideal preparation and measurement. In particular,
$Z_0=1$; $Z_m$ is a normalized contrast, not the survival probability itself.

For an exponential field autocorrelation
$\langle b(t)b(t')\rangle=\sigma^2e^{-|t-t'|/\tc}$, integration over an idle window
$\dd$ yields
\begin{equation}
\label{eq:variance}
 v(\dd)=2\sigma^2\tc^2 f(\dd/\tc),\qquad f(x)=x+e^{-x}-1.
\end{equation}
This expression is shared by Gaussian exponential noise and symmetric telegraph
noise with the same autocorrelation, although their complete phase distributions
differ. The weak-phase single-slot error per Clifford (EPC) is $v/6$.
Its local exponent crosses from
$2$ at $\dd\ll\tc$ to $1$ at $\dd\gg\tc$. For Gaussian phases specifically,
$\EPC=(1-e^{-v/2})/3$ is exact for the averaged single-slot channel. That formula
does not make a multi-slot correlated Gaussian decay a single exponential, and
finite phase saturation can move the observed exponent outside its weak-phase
limits.

In hardware analysis we fit the phenomenological local model
\begin{equation}
\label{eq:gapfit}
 \EPC(\dd)\simeq\varepsilon_0+A\dd^\nu,\qquad
 \nu=\frac{\mathrm{d}\ln(\EPC-\varepsilon_0)}{\mathrm{d}\ln\dd}.
\end{equation}
The offset must be fitted and the abscissa is $\dd$, not gate duration plus $\dd$.
The fit is an effective description on a finite interval, rather than an exact
power law through the full crossover. A quadratic response can arise from a
deterministic frequency error as well as slowly fluctuating noise.

\subsection{An exact reset equivalence}
\label{sec:equivalence}
\begin{proposition}
\label{prop:reset}
Let $b(t)=\pm\sigma$ be a stationary symmetric random telegraph noise (RTN)
process with flip rate
$\gamma=1/(2\tc)$. Each slot has an arbitrary fixed real modulation of its coupling
to this field, and ideal independent Clifford twirling produces the even factor
$p(\phi_k)$. Then
\begin{equation}
\label{eq:reset}
 Z_m=\prod_{k=1}^{m}\lambda_k,\qquad
 \lambda_k=\mathbb{E}[p(\phi_k)].
\end{equation}
The same contrasts result if the environment is independently reset to its stationary
distribution before every slot while preserving each slot's phase law.
\end{proposition}

To prove the statement, define the weighted transition matrix
$[M_k]_{s's}=\mathbb{E}[p(\phi_k)\mathbb{1}_{s_{\rm end}=s'}\mid s_{\rm start}=s]$.
The telegraph dynamics are invariant under global sign reversal, and the weight is
even. Therefore $S M_k S=M_k$, where $S$ swaps the two telegraph states. The
one-dimensional symmetric subspace is invariant, so
$M_k\pi=\lambda_k\pi$ for $\pi=(1,1)^{\mathsf T}/2$. Applying the matrices in order
gives Eq.~\eqref{eq:reset}, including slots with different durations or modulation.
Uncoupled environmental evolution between slots also preserves $\pi$. Independent
stationary resets give the same product directly. $\blacksquare$

For identical slots the result reduces to $Z_m=\lambda^m$. The stronger statement
is that a scan over any set of durations remains identical to a slot-reset model.
This reset model retains correlations within each slot; it is memoryless across
slot boundaries, not necessarily a continuous-time Lindblad semigroup. Thus a gap
scan can reject a white-noise duration law without certifying inter-slot memory.
No procedure using only these ideal mean contrasts distinguishes this particular
pair. Sequence-resolved distributions may contain further information and are
not covered by this no-go statement.

The proof relies on a single symmetric two-state fluctuator and gate-independent
coupling. Multiple fluctuators, unequal switching rates, or gate errors that depend
on the environment need not share this equivalence. An even modulation of the
same scalar phase cannot break it; changing the parity of the measured response can.
The quasi-static equal-branch case is already within the blindness mechanism
described in Ref.~\cite{BlindSpots2026}. The construction above makes the
continuous telegraph/reset pair explicit for arbitrary fixed slot modulations.

\section{Phase cycling recovers the hidden correlator}
\label{sec:cycle}
Apply a known virtual rotation $R_z(s\beta)$ in a slot, with $s=\pm1$, keeping this
rotation out of the Clifford inverse. The twirled factor becomes
\begin{align}
\label{eq:bias}
 p_s(\phi)&=a(\phi)-s b(\phi),\nonumber\\
 a(\phi)&=\frac{1+2\cos\beta\cos\phi}{3},\quad
 b(\phi)=\frac{2\sin\beta\sin\phi}{3}.
\end{align}
For two slots, let $Z_{st}=\mathbb{E}[p_s(\phi_1)p_t(\phi_2)]$, and obtain the
matched marginal responses $z_{k,s}=\mathbb{E}[p_s(\phi_k)]$ from single-slot
experiments. Stationarity or engineered replay must ensure that the single-slot
law agrees with the corresponding marginal of the two-slot experiment.

\begin{proposition}
\label{prop:cycle}
For any classical joint distribution of $(\phi_1,\phi_2)$ obeying the dephasing and
Clifford-independence assumptions above, define
\begin{align}
 J_\beta&=\frac14\sum_{s,t=\pm1}st\,Z_{st},\qquad
 u_k=\frac{z_{k,+}-z_{k,-}}2,\nonumber\\
 C_\beta&=J_\beta-u_1u_2.
\end{align}
Then
\begin{equation}
\label{eq:covariance}
 \boxed{C_\beta=\frac49\sin^2\!\beta\,
           \Cov(\sin\phi_1,\sin\phi_2).}
\end{equation}
\end{proposition}

Substitution of Eq.~\eqref{eq:bias} shows that the sum weighted by $st$ removes
the even and mixed-parity terms, leaving $\mathbb{E}[b(\phi_1)b(\phi_2)]$.
The single-slot differences are $u_k=-\mathbb{E}[b(\phi_k)]$. Subtracting their
product proves Eq.~\eqref{eq:covariance}. $\blacksquare$

No Gaussian or small-angle approximation was used. A nonzero $C_\beta$ excludes
independence of the two phases within this measurement model. A fixed detuning
has zero covariance even when both marginal sine responses are nonzero; omitting
the connected subtraction would fail this essential control. Independent resets
also give zero, whereas a slowly fluctuating sign shared by both slots generally
does not. The choice $\beta=\pi/2$ maximizes the coefficient, but also introduces
a substantial known coherent rotation and is not an ordinary low-error EPC
measurement. In particular, a revival under this controlled bias is not interpreted
as evidence of quantum memory.

The witness is sufficient but incomplete. Correlated distributions can have zero
sine covariance, including phase aliases near integer multiples of $\pi$. Multiple
values of the idle duration or coupling strength can address particular zeros but
do not turn this second-moment witness into a universal dependence test. In the
weak-phase limit, $C_\beta\simeq(4/9)\sin^2\!\beta\,\Cov(\phi_1,\phi_2)$.

\subsection{Closed response for continuous telegraph noise}
\label{sec:response}
For two equal windows of duration $\dd$ separated by an uncoupled interval $h$,
define $\kappa^2=\gamma^2-\sigma^2$. Exact telegraph evolution gives
\begin{equation}
\label{eq:closed}
 C_\beta(h)=\frac49\sin^2\!\beta\,e^{-h/\tc}
 \left[\sigma e^{-\gamma\dd}\frac{\sinh(\kappa\dd)}{\kappa}\right]^2.
\end{equation}
For imaginary $\kappa$, the bracket is evaluated with a real sine; at $\kappa=0$
its continuous limit is $\sigma\dd e^{-\gamma\dd}$. The transfer derivation is
given in Appendix~\ref{app:transfer}. In the quasi-static limit, Eq.~\eqref{eq:closed}
becomes $(4/9)\sin^2\!\beta\sin^2(\sigma\dd)$. At fixed window duration and
nonzero response, its dependence on an ideal uncoupled separation directly gives
$\tc^{-1}=-\mathrm{d}\ln C_\beta/\mathrm{d}h$.

This last relation is a model prediction, not a claimed hardware correlation-time
measurement. A bare delay on a physical qubit usually remains coupled to its
environment; engineering an effectively uncoupled separation requires a separate
control and its error characterization. The injected hardware experiment below
tests the shared-sign limit and an independent reset, while finite switching times
and the $h$ dependence are validated numerically.

\section{Estimation and finite-sample scope}
\label{sec:statistics}
One block contains the four two-slot settings $++$, $+-$, $-+$, $--$,
followed by marginal settings $1+$, $1-$, $2+$, $2-$. These share a
Clifford pair and a programmed phase pair. Let $P_{i,j}$ be the observed survival
fraction in block $i$ and setting $j$. The block quantities are
\begin{align}
 J_i&=(P_{i,++}-P_{i,+-}-P_{i,-+}+P_{i,--})/2,\nonumber\\
 U_i&=P_{i,1+}-P_{i,1-},\quad V_i=P_{i,2+}-P_{i,2-}.
\end{align}
Using distinct blocks for the product of means gives the unbiased estimator
\begin{equation}
\label{eq:ustat}
 \widehat C=\overline J-
 \frac{(\sum_i U_i)(\sum_i V_i)-\sum_i U_iV_i}{n(n-1)}.
\end{equation}
Independence across blocks makes each $i\ne j$ product unbiased for
$\mathbb{E}U\mathbb{E}V$, while arbitrary dependence within a block is allowed.
This also removes the finite-sample covariance bias of simply multiplying two
paired sample means.

The practical interval resamples whole blocks, keeping phase settings and matched
control arms together. A percentile bootstrap is an empirical approximation,
not a finite-sample coverage theorem. We assess it on independently seeded
simulation trials using a fixed decision rule: a two-sided nominal 95\% interval
must exclude zero. No threshold is optimized on these validation trials.

A separate conservative confidence set is available because $J_i,U_i,V_i\in[-1,1]$.
For independent identically distributed blocks, Hoeffding's inequality and a union
bound imply that all three population means lie within
\begin{equation}
\label{eq:radius}
 \epsilon_n=\sqrt{2\ln(6/\alpha)/n}
\end{equation}
of their sample means with probability at least $1-\alpha$. Intersect each mean
interval with $[-1,1]$, then evaluate the minimum and maximum of $j-uv$ over
that rectangular set. The resulting interval covers $C_\beta$ at least
$1-\alpha$ and need not be centered on Eq.~\eqref{eq:ustat}. Its conservatism
is reported separately from bootstrap performance.

Independent blocks are a substantive assumption. Repeated observations of a
slowly drifting native environment can violate it. Such observations require
time ordering, an appropriate block length or independent acquisitions; the
independent injected trajectories used in simulation do not establish coverage
for arbitrary hardware drift.

For an affine readout response $P_{\rm obs}=aP+b$, phase differences cancel $b$.
The joint term scales as $a$ and the product of marginal differences as $a^2$.
We therefore calibrate $a$ from prepared zero and one states and divide each of
$J,U,V$ by the measured gain before Eq.~\eqref{eq:ustat}. Hardware bootstrap
replicates resample those calibration counts as well. The finite bound above
applies to the ideal bounded observations; it is not asserted unchanged after
estimating and dividing by readout gain. Stable, setting-independent state
preparation and measurement (SPAM) and
gate errors remain experimental assumptions tested in part by the controls.

\section{Numerical validation}
\label{sec:simulation}
The exact transfer calculation, closed response, and explicit random-Clifford
Bloch propagation are implemented separately. A scan over 19 logarithmically
spaced correlation times and three idle durations checks reset equivalence for
lengths $m=0,\ldots,50$. The maximum discrepancy is $\cycleResetError$.
The closed form agrees with the eight transfer-matrix measurement means to
$\cycleFormulaError$, including a separate critical-damping check. Enumeration
of all $24^2$ Clifford pairs agrees with the product-twirl identity to
$\cycleTwirlError$ for both weak and strong test angles.

\begin{figure*}[t]
\includegraphics[width=\textwidth]{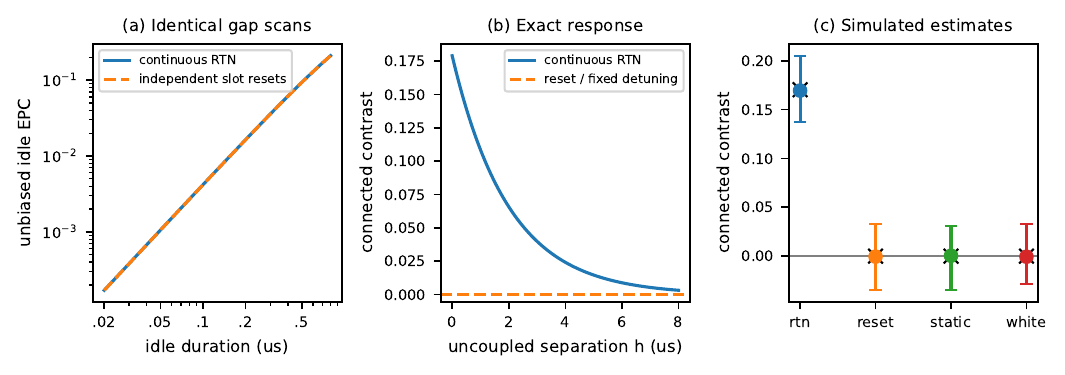}
\caption{\label{fig:phasecycle}Simulation and exact theory. (a) Continuous
telegraph noise and independent slot resets have identical mean gap scans.
(b) The phase-cycle covariance distinguishes them; the separation dependence is
the ideal uncoupled-interval prediction of Eq.~\eqref{eq:closed}.
(c) Median and central 95\% sampling ranges over independent simulated experiments,
with crosses marking population values. These are sampling-distribution ranges,
not confidence intervals from one experiment.}
\end{figure*}

Figure~\ref{fig:phasecycle} displays the exact response and sampling distributions.
The sampling study uses \cycleTrials{} independent trials per class,
\cycleBlocks{} independent blocks per trial, and 64 binomial shots per setting.
The fixed continuous-RTN parameters are $\sigma=1.6$ rad/$\mu$s,
$\tc=2$ $\mu$s, $\dd=0.5$ $\mu$s and $h=0.1$ $\mu$s.
Controls comprise independent resets with the same complete single-slot phase
law, a deterministic phase, and independent Gaussian slot phases. The exact
RTN covariance is \cycleTruth{}. Results are given in Table~\ref{tab:validation};
coverage variation makes the distinction between empirical and guaranteed
intervals material.

\begin{table}[t]
\caption{\label{tab:validation}Held-out simulation calibration. Entries are
percentages over \cycleTrials{} trials per class. Detection means the nominal
two-sided bootstrap interval excludes zero. The conservative bound uses
Eq.~\eqref{eq:radius}.}
\begin{ruledtabular}
\begin{tabular}{lrrr}
Model & Detection & Bootstrap & Finite bound \\
 & & coverage & coverage \\
\colrule
Continuous RTN & \cycleRtnDetect & \cycleRtnCoverage & \cycleRtnFinite \\
Independent reset & \cycleResetDetect & \cycleResetCoverage & \cycleResetFinite \\
Fixed detuning & \cycleStaticDetect & \cycleStaticCoverage & \cycleStaticFinite \\
Independent Gaussian & \cycleWhiteDetect & \cycleWhiteCoverage & \cycleWhiteFinite \\
\end{tabular}
\end{ruledtabular}
\end{table}

\subsection{Additional assumption checks}
An additional 200 randomly generated schedules use unequal slots with piecewise real modulation of either sign. Their continuous/reset mean discrepancy is below $10^{-14}$. Testing 100 discrete, nonsymmetric joint phase laws and varying the cycle angle verifies Eq.~\eqref{eq:covariance} below the same tolerance. A deliberately asymmetric telegraph example gives a reset discrepancy 0.000852; the symmetry assumption cannot simply be dropped.

\subsection{An ideal resource-matched Ramsey comparator}
\label{sec:ramsey}
A coherent Ramsey/echo pair measures the mean cosines of the sum and difference of two phases. Two single-slot $Y$-quadrature settings measure their sine means. Thus a four-setting ideal construction gives
\begin{align}
 C_R={}&\tfrac12\mathbb{E}[\cos(\phi_1-\phi_2)-\cos(\phi_1+\phi_2)]\nonumber\\
 &-\mathbb{E}[\sin\phi_1]\mathbb{E}[\sin\phi_2]\nonumber\\
 ={}&\Cov(\sin\phi_1,\sin\phi_2).
\end{align}
Independent state-amplitude propagation verifies all four ideal responses. This comparator is an ideal Ramsey/echo construction motivated by established correlation spectroscopy~\cite{Yan2012,Zohar2026}; it is not a reproduction of the RESOLUTE experiment or its full control sequence.

We simulate 200 trials per class, with 128 independent phase-pair blocks. RB uses eight settings and 8 shots per setting; Ramsey uses four settings and 16 shots. Both therefore use 8,192 circuit shots and 12,288 sensing-window exposures per trial. They share the same sampled phase pairs and use the distinct-block connected estimator. RB estimates are multiplied by $9/4$ so both target $C_R$. Noise parameters match the preceding validation; shots and blocks differ. This compares shot and sensing-window resources, not gate duration or wall time.

\begin{table}[t]
\caption{\label{tab:ramsey}Root mean squared error for the same sine covariance in ideal simulation. Neither method has native drift or imperfect gates in this comparison.}
\begin{ruledtabular}
\begin{tabular}{lrrr}
Model & RB & Ramsey/echo & Ratio \\
\colrule
Continuous RTN & 0.0777 & 0.0269 & 2.89 \\
Independent reset & 0.0798 & 0.0432 & 1.85 \\
Fixed detuning & 0.0793 & 0.0188 & 4.22 \\
Independent Gaussian & 0.0824 & 0.0318 & 2.59 \\
\end{tabular}
\end{ruledtabular}
\end{table}

The Ramsey/echo estimate is more precise in all tested classes (Table~\ref{tab:ramsey}). The supplementary data retain all trial estimates and paired Monte Carlo uncertainty for each ratio. These results support the RB embedding as an interpretive construction, not an efficiency claim.

\section{Hardware controls}
\label{sec:hardware}
\subsection{A test that holds the slot marginals fixed}
The phase-cycle design is saved before submission and uses a shared Clifford pair
across four arms. With $\theta=0.8$ rad, the correlated arm draws a random sign
and injects $(\phi_1,\phi_2)=(s\theta,s\theta)$. The reset arm draws two independent
signs, yielding $(s\theta,t\theta)$ with identical single-slot distributions.
The deterministic arm injects $(\theta,\theta)$, and the last arm has no injected
phase. The ideal connected response is $(4/9)\sin^2\theta$ for the correlated arm
and zero for each control. In particular, ordinary unbiased RB means coincide
for the first three arms because $\cos(\theta)=\cos(-\theta)$.

Each block contains eight settings. Each slot contains a 320 ns idle; the injected phase and
the $\pm\pi/2$ bias are virtual rotations protected by circuit barriers.
Circuits use Qiskit~\cite{Qiskit2024}. The programmed rotations remain outside the Clifford inverse. Readout calibration
adds 64 circuits for each prepared state. The processor is \posBackend{} q46.
Initial submissions used individually serialized circuits. A subsequent version
groups phase bindings into shared Clifford templates without changing their
unitaries; the template order and the order of settings within each template are
randomized. Sample counts and acquisition outcomes are reported below.
Randomization reduces order confounding but does not itself prove stationarity.

The primary comparison, chosen before collection, is the paired difference between
the correlated and reset contrasts. Per-arm intervals additionally check the fixed
detuning and no-injection controls. Injected phases are held fixed over each
circuit's shot group; native fluctuations are not replayed between distinct circuits.
Thus the experiment calibrates recovery of engineered inter-slot dependence rather
than detecting a native hardware memory process.

\subsection{Injected phase-cycle results}
The archive contains 2 completed acquisitions.
Acquisition 1 uses 128 programmed blocks per arm and 8 shots per setting (4,224 circuit evaluations, 33,792 circuit shots), with random seed 2026090501 and readout gain 0.9980.
Acquisition 2 uses 128 programmed blocks per arm and 8 shots per setting (4,224 circuit evaluations, 33,792 circuit shots), with random seed 2026090502 and readout gain 0.9883.
The second acquisition was specified after inspecting the first result, with a fresh seed and the same block count, shots, phase amplitude, observable and controls. Both acquisitions used the same qubit on the same day; they test within-session repeatability, not robustness across devices or calibration cycles.
Intervals use 4,000 paired block bootstrap replicates and resample readout calibration counts. They are empirical intervals conditional on the stability assumptions in Sec.~\ref{sec:statistics}. The conservative coverage guarantee for bounded uncorrected observations is not transferred to these calibration-corrected hardware intervals.

\begin{figure}[t]
\includegraphics[width=\columnwidth]{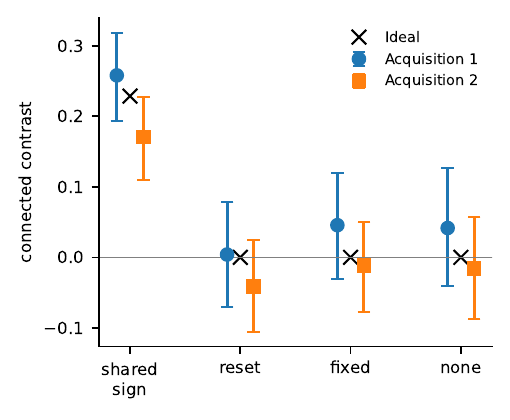}
\caption{\label{fig:cyclehardware}Hardware response to engineered phase correlations. Points and error bars are connected estimates and paired block bootstrap intervals for each acquisition; crosses are ideal population predictions.}
\end{figure}

The primary correlated-minus-reset contrast in acquisition 1 is $0.2540$, with nominal 95\% interval $[0.1775, 0.3310]$. It excludes zero in the predicted direction.
The primary correlated-minus-reset contrast in acquisition 2 is $0.2114$, with nominal 95\% interval $[0.1359, 0.2851]$. It excludes zero in the predicted direction.
All individual negative-control intervals contain zero.
In acquisition 2, the shared-sign interval does not contain the ideal population prediction $0.2287$. The positive separation therefore does not imply exact quantitative agreement with an ideal device.
All arms appear in Table~\ref{tab:cyclehardware} and Fig.~\ref{fig:cyclehardware}.

\begin{table}[t]
\caption{\label{tab:cyclehardware}Measured injected phase-cycle contrasts and empirical 95\% intervals. The ideal population contrast is $0.2287$ for shared signs and zero for the other arms.}
\begin{ruledtabular}
\begin{tabular}{lcc}
Arm & Estimate & 95\% interval \\
\colrule
\multicolumn{3}{c}{Acquisition 1} \\
Shared sign & 0.2580 & $[0.1935, 0.3182]$ \\
Independent reset & 0.0040 & $[-0.0704, 0.0779]$ \\
Fixed detuning & 0.0456 & $[-0.0304, 0.1195]$ \\
No injection & 0.0416 & $[-0.0410, 0.1265]$ \\
\colrule
\multicolumn{3}{c}{Acquisition 2} \\
Shared sign & 0.1704 & $[0.1091, 0.2272]$ \\
Independent reset & -0.0409 & $[-0.1053, 0.0249]$ \\
Fixed detuning & -0.0119 & $[-0.0769, 0.0495]$ \\
No injection & -0.0154 & $[-0.0874, 0.0568]$ \\
\end{tabular}
\end{ruledtabular}
\end{table}

The archive also retains 3 unsuccessful jobs with no usable counts and 45 seconds of recorded charges. They are excluded from the estimator and not counted as replications. Reductions in sample count and Runtime packaging preceded the first successful measurement. The completed results are engineered-noise proofs of concept; they establish neither native memory nor a physical correlation-time scan.

\subsection{Readout calibration sensitivity}
A zero-error calibration count is not evidence of a zero population error rate. To assess this boundary, we construct two exact binomial intervals at 97.5\% per acquisition and combine them into a simultaneous 95\% interval for the readout gain by a union bound. With experimental survivals held fixed, we recompute the primary contrast over a dense gain grid.
For acquisition 1, this calibration-only range is $[0.2536, 0.2590]$.
For acquisition 2, this calibration-only range is $[0.2093, 0.2171]$.
These ranges do not include experimental sampling error and are not confidence intervals for the primary contrast. They indicate that the observed positive point estimates do not depend on setting unobserved calibration errors exactly to zero. They do not test drift or setting-dependent SPAM.

\subsection{Complementary idle-gap calibration}
The earlier campaign used synthetic dephasing trajectories with known correlation
times in idle windows. Effective exponents on \posBackend{} q\posQubit{} were
$\nu=\nuMeasFast\pm\nuErrFast$, $\nuMeasMid\pm\nuErrMid$, and
$\nuMeasQs\pm\nuErrQs$ for fast, crossover and quasi-static arms, respectively.
The largest discrepancy from the stored response prediction was \posMaxZ{}
standard errors. These are calibration observations for Eq.~\eqref{eq:gapfit},
not independent evidence establishing priority or universal classifier performance.

Several limitations affect that legacy analysis. The baseline-subtracted EPC is
a leading-order approximation: even independent depolarizing contractions compose
multiplicatively, so $r_{\rm combined}=r_0+r_i-2r_0r_i$ and subtraction does not
cancel background exactly. The stored analysis uses data-dependent signal-to-noise
cuts and separately estimated EPC uncertainties rather than a complete paired
joint fit. Its interval coverage under those choices has not been established.
The quasi-static Hahn control demonstrates refocusing of the programmed phase;
because a slot-integrated phase is distributed proportionally between echo
segments, it does not validate the full within-slot switching response of finite-time
telegraph noise. These observations motivate the more direct matched-marginal
phase-cycle test.

Repeated-sequence variance removes a separate confound: different Clifford
sequences can have different intrinsic fidelities even with stationary noise.
The injected quasi-static arm gave $\Wrep=\wvarRepQsDD$ and native controls near
unity. Nevertheless, this statistic probes changes between repeated circuit
executions, a different timescale from memory between gates. Shot groups that
average many independent trajectories may suppress it. It is a diagnostic, not a
necessary condition for all inter-gate memory.

\section{Native survey and limits of a quantitative null}
\label{sec:null}
The earlier screen covered \totalQubits{} qubits across \posBackend{} and
\devTwo{}. Two gap-exponent candidates did not reproduce on follow-up.
On \posBackend{}, q53 changed from $1.87\pm0.22$ to $1.26\pm0.14$.
On \devTwo{}, q\devTwoTopQubit{} changed from
$\devTwoTopNu\pm\devTwoTopErr$ to $\devTwoConfNu\pm\devTwoConfErr$.
These measurements support the statement that there was no replicated detection
under the screening rule. Regression after selecting a maximum is compatible
with selection effects; temporal drift is another explanation, and the data do
not uniquely attribute the changes to either cause.

A stored six-gap EPC curve on q\nullQubit{} has the effective exponent
$\nullNu\pm\nullNuErr$. To ask whether it supports an exclusion, consider
\begin{equation}
\label{eq:mixture}
 r(\dd)=\varepsilon_0+a\left[(1-f)\frac{\dd}{\dd_{\max}}
       +f\frac{g(\dd/\tc)}{g(\dd_{\max}/\tc)}\right],
\end{equation}
where $g(x)=x+e^{-x}-1$, $a\ge0$, $\varepsilon_0\ge0$, and $0\le f\le1$.
The fraction is defined at the largest gap. At every proposed $(f,\tc)$, we refit
both nuisance parameters by weighted nonnegative least squares. Holding them
at their previously fitted values understates their uncertainty.

With the stored diagonal EPC error model, the minimum residual $\chi^2$ ranges
from \auditChiMin{} to \auditChiMax{} over the 13 tested correlation times
20--200,000 ns. All exceed \auditBallCut{}, the 95th percentile of $\chi^2_6$.
If the six-dimensional Gaussian covariance were known, the ellipsoid
$(y-\mu)^{\mathsf T}\Sigma^{-1}(y-\mu)\le\chi^2_{6,0.95}$ would cover the true
mean with probability 95\%. An empty intersection with the proposed model
family on this grid signals model incompatibility, rather than exclusion of
every possible correlated contribution. Here the covariance is itself estimated
and omits unavailable cross-gap covariance, so this is an adequacy diagnostic.

\begin{figure}[t]
\includegraphics[width=\columnwidth]{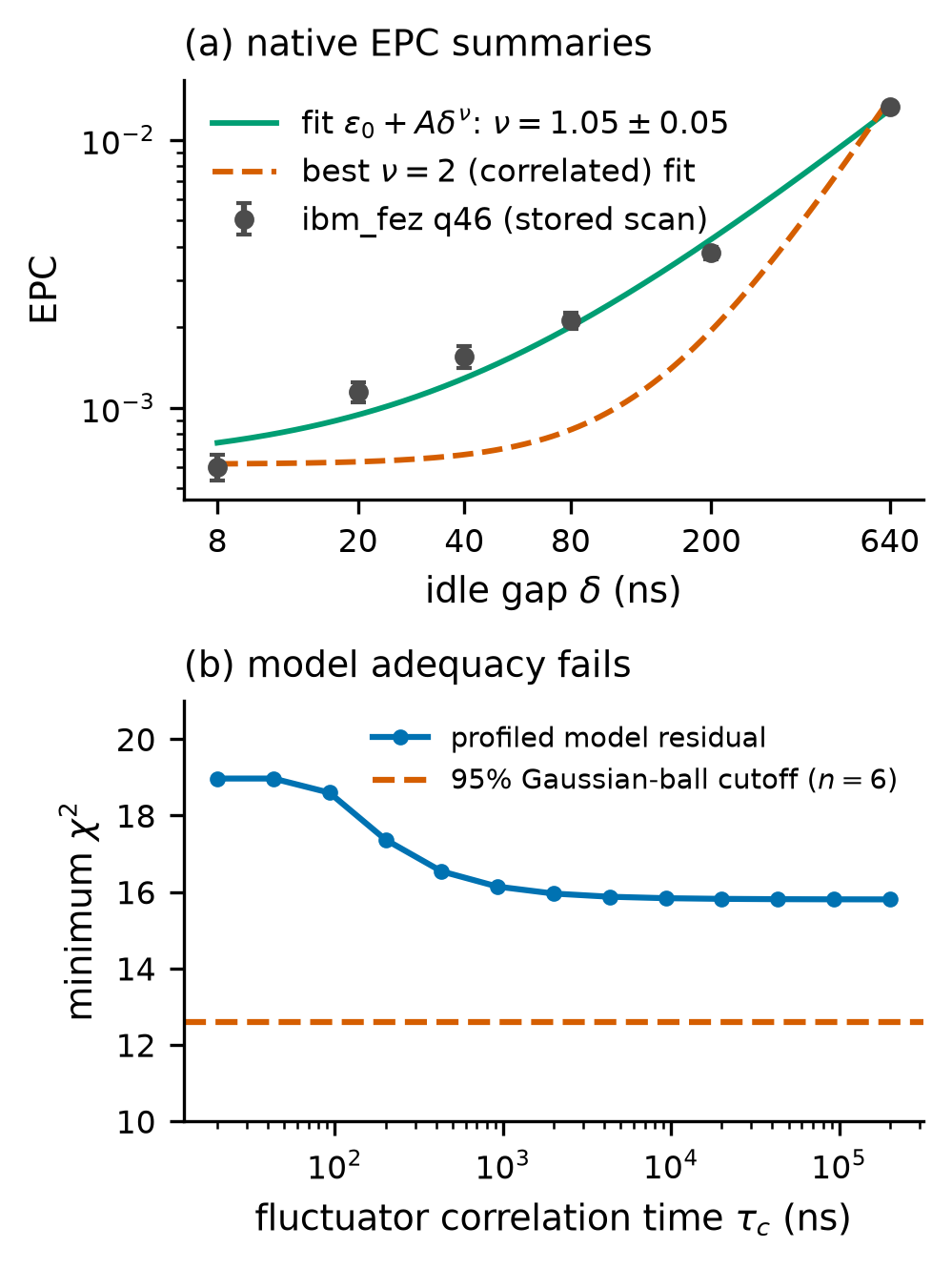}
\caption{\label{fig:audit}Audit of the stored native EPC summaries.
(a) Offset-aware phenomenological fit. (b) Best residual after profiling
nonnegative nuisance parameters, versus the Gaussian-ball cutoff.
The tested model curves fail this adequacy diagnostic, so no exclusion region
is shaded or claimed.}
\end{figure}

Consequently, we do not quote an unconditional 95\% correlated-noise exclusion
from these summaries. A profile-likelihood threshold of $\Delta\chi^2=2.71$
does not repair a poor absolute fit, establish finite-sample coverage, or provide
a simultaneous bound over correlation times. A quantitative exclusion requires
reanalysis of raw blocks with covariance, checks of model adequacy, and coverage
calibration for the resulting procedure. Figure~\ref{fig:audit} displays this
distinction explicitly.

\section{Discussion}
The reset equivalence explains why a useful duration-dependent noise measurement
can still leave inter-slot dependence unidentified. Phase cycling changes the
observable rather than merely improving the precision of the same even response.
The connected subtraction is essential: it separates shared random phase from
a deterministic frequency shift, while the reset control tests specificity at
fixed single-slot noise statistics.

The current scope is classical idle dephasing with Clifford-independent phases.
Finite gate durations, gate-dependent error, leakage, nonstationary SPAM and
unmatched single-slot marginals can all invalidate the simple interpretation.
The fixed-detuning and no-injection controls test some of these effects but
cannot establish their absence under all operating conditions. The exact
telegraph formula applies to ideal uncoupled separation; demonstrating a
physical lag scan requires additional control validation.

Only one- and two-slot circuits are needed, but multiple settings and independent
blocks are required. The ideal resource-matched comparison in
Sec.~\ref{sec:ramsey} favors Ramsey/echo estimation in every tested noise class.
The present method therefore offers an explicit connection to the Clifford RB
observable, rather than a demonstrated precision advantage. It does not inherit
ordinary RB's fitted separation of gate decay and arbitrary SPAM coefficients.
Repeated acquisitions across calibration epochs and more general gate-error
models are needed to assess its experimental generality.

The finite-sample construction also separates what is proved from what is measured.
The conservative mean-rectangle confidence set has a coverage statement under
independent bounded blocks. The bootstrap is more precise but its coverage is
empirical. Neither implies quantum-memory certification or a worst-case gate-error
bound. Similarly, failure to replicate a native screening candidate is a useful
negative result without implying an accurately calibrated exclusion fraction.

\section{Conclusion}
For a symmetric telegraph environment, mean gap-scan RB remains exactly unchanged
when memory across slot boundaries is erased. A controlled phase cycle recovers
the sine-phase covariance that the ordinary even Clifford response hides.
The exact identity supplies explicit independent-noise and deterministic-detuning
nulls, a telegraph response function, and a connected estimator with separately
stated empirical and conservative uncertainty assessments. The hardware controls
and the native-data audit distinguish an operationally testable memory observable
from stronger claims that the present data do not support.

\section*{Data and code availability}
The accompanying repository contains the data, analysis code and
reproduction instructions. Its inventory distinguishes observed hardware
counts, simulated data, failed acquisitions and legacy summaries; four
older datasets remain summary-only. Code uses a custom source-available
license requiring attribution in public work substantially using it, and
original data and figures use CC BY 4.0. See the repository attribution
and citation files for the authors and license scope.
The companion project repository is \href{https://github.com/Mirza-Samad-Ahmed-Baig/quantum-phase-cycled-rb}{\texttt{quantum-phase-cycled-rb}}.

\begin{acknowledgments}
We acknowledge the use of IBM Quantum services. The views expressed are those
of the authors and do not reflect the official policy or position of IBM or
the IBM Quantum team.
\end{acknowledgments}

\appendix
\section{Telegraph response derivation}
\label{app:transfer}
Let
\begin{equation}
 Q=\begin{pmatrix}-\gamma&\gamma\\\gamma&-\gamma\end{pmatrix},\qquad
 D=\operatorname{diag}(\sigma,-\sigma).
\end{equation}
Feynman--Kac propagation of a phase characteristic function over an idle window
is $E_k=\exp[(Q+ikD)\dd]$, for $k=-1,0,1$. The biased one-slot weighted matrix is
\begin{equation}
 M_\beta=e^{Qh}\frac{E_0+e^{i\beta}E_1+e^{-i\beta}E_{-1}}3.
\end{equation}
For equal biases, $Z_m=\boldsymbol{1}^{\mathsf T}M_\beta^m\pi$; distinct
phase signs are obtained by ordered multiplication of the corresponding matrices.
At zero bias the symmetric stationary subspace is invariant. The sine component
$S_\dd=(E_1-E_{-1})/(2i)$ instead exchanges the symmetric and antisymmetric
subspaces. In their normalized basis its off-diagonal entry is
\begin{equation}
 R=\sigma e^{-\gamma\dd}\frac{\sinh(\kappa\dd)}{\kappa}.
\end{equation}
The antisymmetric component decays by $e^{-2\gamma h}$ between windows.
Stationary reflection symmetry gives $\mathbb{E}\sin\phi_k=0$, while
$\mathbb{E}[\sin\phi_1\sin\phi_2]=R^2e^{-2\gamma h}$.
Substitution into Eq.~\eqref{eq:covariance} proves Eq.~\eqref{eq:closed}.

\bibliographystyle{apsrev4-2}
\bibliography{refs}
\end{document}